\documentstyle[prl,aps,multicol,psfig]{revtex}
\begin{document}
\draft

\title{Scaling in DNA unzipping models: denaturated loops and end-segments as
branches of a block copolymer network}
\author{Marco Baiesi$^1$, Enrico Carlon$^{1,2}$, and Attilio L.~Stella$^{1,3}$}
\address{
$^1$INFM - Dipartimento di Fisica, Universit\`a di Padova,
I-35131 Padova, Italy\\
$^2$Theoretische Physik, Universit\"at des Saarlandes, D-66123 Saarbr\"ucken, 
Germany\\
$^3$Sezione INFN, Universit\`a di Padova, I-35131 Padova, Italy
}
\date{\today}
\maketitle

\begin{abstract}
For a model of DNA denaturation, exponents describing the
distributions of denaturated loops and unzipped end-segments are determined 
by exact enumeration and by Monte Carlo simulations in two and three
dimensions. The loop distributions are consistent with first order thermal 
denaturation in both cases.
Results for end-segments show a coexistence of two distinct
power laws in the relative distributions, which is not foreseen
by a recent approach in which DNA is treated as a homogeneous network 
of linear polymer segments. This unexpected feature, and the 
discrepancies with such an approach, are explained 
in terms of a refined scaling picture in which a precise distinction 
is made between network branches representing 
single stranded and effective double stranded segments.
\end{abstract}

\pacs{PACS numbers: 05.20.-y,  05.70.Fh, 64.60.-i, 87.14.Gg}

\begin{multicols}{2} \narrowtext

\section{Introduction}

A DNA molecule may undergo transitions from a double stranded to a single 
stranded state, either under the effect of an increase in temperature $T$ 
(thermal denaturation), or through applied forces at one end of the chain 
(mechanical unzipping) \cite{Pola66,Fish66,mechzip}.
In the characterization of such transitions, and in the determination of 
their universal, asymptotic features, substantial progresses were made 
recently by applications of models and methods of polymer 
statistics \cite{mu,All,Kafr00,Caus00,Carl01}. 
Among these progresses is an extension \cite{Kafr00} of the classical 
Poland and Sheraga (PS) model \cite{Pola66}.
In the PS model the partition function of a DNA chain is approximated by
that of a sequence of non-interacting double-stranded segments and denaturated
loops, and the thermal denaturation transition results of second order type
\cite{Pola66,Fish66}. Recently, excluded volume effects between a loop and
the rest of the chain were included in the PS description in an approximate
way \cite{Kafr00}, using results from the theory of polymer networks 
\cite{Dupl86}. This approach predicts a first order denaturation, in agreement 
with very recent numerical studies of models taking fully into account the 
self and mutual avoidance among loops and double segments \cite{Caus00,Carl01}. 
Quite remarkably, the approximate scheme of Ref.~\cite{Kafr00} yields
results which are in good quantitative agreement with 
Monte Carlo simulations \cite{Carl01}. Most recently,
predictions based on the theory of polymer networks were also made for 
the case of mechanical unzipping \cite{Kafr01}.

Besides confirming the expected first order character of thermal 
denaturation, the results of Ref.~\cite{Carl01} demonstrated that
excluded volume effects alone are responsible for this character,
while the difference in stiffness between double and single stranded
DNA, and sequence heterogeneity do not affect the
asymptotic nature of the transition.

All these results rise interesting  and debated ~\cite{comments}
issues and open new perspectives in the 
field. First of all, one would like to test numerically the existing 
analytical estimates, in particular the new ones pertaining to the case of 
mechanical unzipping. The validity of crucial predictions relies on such 
tests, which should also reveal whether the quantitative
success in the case of thermal 
unzipping is rather fortuitous, or there is some deep and systematic basis 
for it. The network picture proposed in Ref.~\cite{Kafr00}, besides allowing 
some elegant and successful estimates, could constitute an important step 
forward in our way of representing the physics of DNA at denaturation.
Progress in the assessment of the validity and limitations of a network 
picture for denaturating DNA
should be allowed by a more careful and systematic analysis of numerical
results for specific models. The possible extension and improvement of 
previous analyses for the relevant three dimensional case is also a main 
motivation of the present work.

In this article we consider a lattice model of DNA both in two and in three 
dimensions ($d$) with excluded volume effects fully implemented. By various 
numerical methods we estimate length distributions for denaturated loops and 
unzipped end-segments. 
In particular,
for the transition in three dimensions we give here exponent estimates 
which extend and improve the results of a previous study \cite{Carl01}. 
Moreover, our results allow to address the basic issues mentioned above,
and to clearly identify some qualitative and quantitative limitations of the 
picture proposed in Ref.~\cite{Kafr00} . The analysis gives also hints which 
allow us to propose a generalization of the polymer network representation
of denaturating DNA.
Our model in two dimensions, besides deserving some interest
in connection with problems like the unzipping of double stranded polymers
adsorbed on a substrate,
offers an ideal context in which to compare the predictions of 
Refs.~\cite{Kafr00,Kafr01}
with numerical results. Indeed, for $d=2$ those predictions are 
based on exactly known network exponents, while in $d=3$ the same exponents 
have been approximately determined.

The model studied here was introduced in Ref.~\cite{Caus00} and 
further analyzed and extended in Ref.~\cite{Carl01}.
We consider two self-avoiding walks (SAWs) of length $N$ on square and cubic 
lattices, described by the vectors identifying the positions of each monomer 
$\vec{r}_1(i)$ and $\vec{r}_2(i)$, with $0 \leq i \leq N$. The walks,
which represent the two DNA strands, have a 
common origin $\vec{r}_1(0) = \vec{r}_2(0)$ and only monomers with the same 
coordinate along the SAW's can overlap each other (i.e. $\vec{r}_1(j)
= \vec{r}_2(k)$ only if $j=k$). 
An overlap corresponds to a bound state of complementary 
DNA base pairs, to which we assign an energy $\varepsilon = -1$, 
thus neglecting the effects of sequence heterogeneities. 
This is an acceptable approximation in view of the fact 
that, at coarse grained level, each monomer 
(site visited by the SAW)
should represent a whole persistence length of the single
strand, which includes several ($\approx 10$) bases.
A denaturated loop of length $l$ occurs whenever, for some $i$,
$\vec{r}_2(i)=\vec{r}_1(i)$, $\vec{r}_2(i+l+1)=\vec{r}_1(i+l+1)$,
and $\vec{r}_2(i+k)\neq \vec{r}_1(i+k)$, $k=1,2,..,l$.

At a temperature $T$ each configuration 
$\omega$ of the two strands appears in the statistics
with probability proportional to its Boltzmann's weight 
$\exp\{-\beta H(\omega)\}$,
where $\beta=1/T$ (Boltzmann's constant $=1$)
and $H$ is $\varepsilon$ times the number of bound base pairs in $\omega$.
The resulting behavior of the DNA model resembles the scaling of 
a $2N$-step SAW as long as the
inverse temperature $\beta$ is lower than a critical value $\beta_c$.
For $\beta>\beta_c$ the strands are typically 
paired in a sequence of bound segments
alternating with denatured loops, the latter gradually shrinking 
and becoming rarer for increasing $\beta$.
For $\beta>\beta_c$ and $N\to\infty$ the scaling of an $N$-step SAW 
holds. Around $\beta_c$ there is a crossover from one regime to the other, and
exactly at $\beta_c$ peculiar scaling behaviors are expected
for the distributions of loops and end-segments.

\section{The two dimensional case}

\begin{figure}[b]
\centerline{
\psfig{file=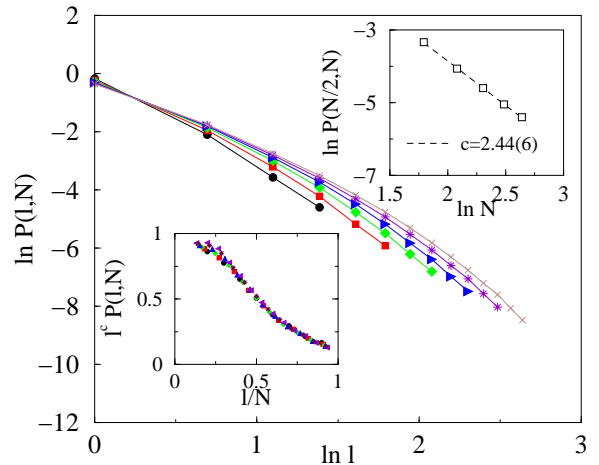,height=6.4cm}}
\vskip 0.2truecm
\caption{$\ln P(l,N)$ vs.\ $\ln l$ for various $N$ values. 
Upper inset: $\ln P(N/2,N)$ vs.\ $\ln N$; data are well-fitted 
by a line from which we estimate $c=2.44(6)$. 
Lower inset: Scaling collapse of the data.}
\label{fig:01}
\end{figure}

We focus first on the two-dimensional case. 
Fig.~\ref{fig:01} plots the logarithm of $P(l,N)$, the pdf
 of finding a denaturated loop of length $l$ within a chain of total 
length $N$, as a function of $\ln l$ at the transition point. The
behavior of this pdf at denaturation determines the order of the transition
\cite{Pola66,Kafr00,Carl01}. The data are obtained from exact enumerations for walks up 
to $N=15$. For each chain length, the pdf is sampled at the temperature
corresponding to the specific heat maximum. This choice offers the advantage
that, if the specific heat diverges at the transition, like in our case,
the correct transition temperature is asymptotically singled out for
$N \to \infty$. Thus, in this limit, the sequence of distributions should 
automatically approach the critical pdf, for which we expect power-law scaling. 

\begin{figure}[b]
\centerline{
\psfig{file=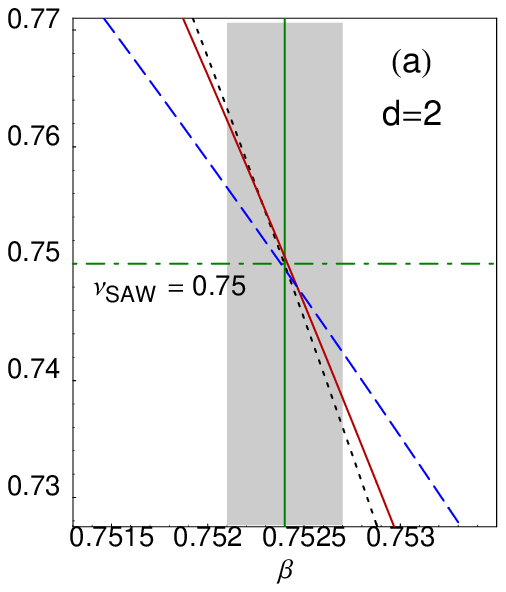,width=4.0cm}
\psfig{file=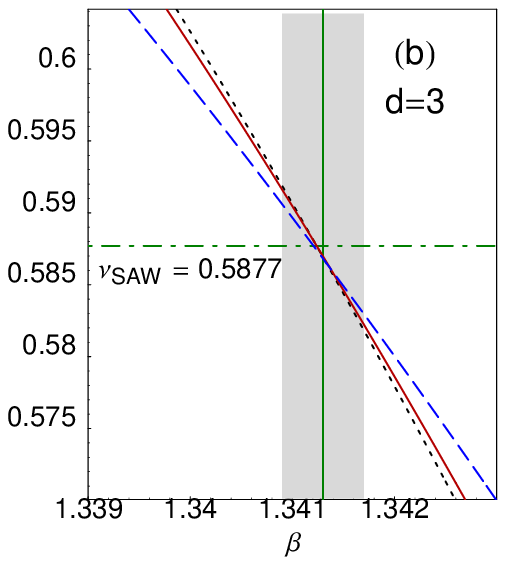,width=4.15cm}
}
\vskip 0.2truecm
\caption{(a) Effective exponent $\nu$ of the end-to-end distance, for 
$N\in A=\{80,\,120,\,160,\,240\}$ (dashed curve), 
$N\in B=\{160,\,240,\,320,\,480\}$ (dotted curve), and  
$N\in A \cup B$ (continuous curve),
as a function of $\beta$ in $d=2$. 
The horizontal dotted line marks the exactly known SAW $\nu=3/4$,
while the vertical line is our estimate $\beta_c = 0.7525(3)$ (the error
is indicated by the gray band).
(b) Similar plots for $d=3$ 
 ($A=\{80,\,120,\,160,\,240\},\, B=\{120,\,160,\,240,\,320\}$). 
In this case the intersections
are even better localized around the expected SAW 
 $\nu\approx 0.5877$. The transition temperature determined
in this way is almost coincident with the estimate 
$\beta_c=1.3413(4)$  \protect\cite{Caus00}
indicated by the vertical line and the gray band.}
\label{fig:02}
\end{figure}

Finite-size effects should be described by $P(l,N) = l^{-c} q(l/N)$,
with $q$ a suitable scaling function. In order to determine $c$,
we consider, e.g., $P(N/2,N)$, which should scale 
$\propto N^{-c}$ for $N \to \infty$. 
Such quantity is shown in the upper inset 
of Fig.~\ref{fig:01}, plotted as a function of $N$ in a log-log scale. From a 
linear fit of the data we obtain the estimate $c=2.44(6)$. The lower inset 
shows a scaling collapse of the data, obtained with $c=2.44$. 
The good quality of the 
collapse indicates that the system is very close to the asymptotic regime
although the chains are quite short.
We also performed a Monte Carlo determination of $c$ using the 
pruned enriched Rosenbluth Method 
(PERM) \cite{perm}, through which walks are generated by a growth procedure.
In this case the critical temperature at which the pdf was sampled
was determined by carefully monitoring the scaling with $N$ of the
average end-to-end
distance $\langle|\vec{r}_2(N)- \vec{r}_1(N)|\rangle$ as a function of
temperature. The effective exponents describing the growth
with $N$ of this quantity are plotted as a function of $\beta$
for different chain lengths in Fig.~\ref{fig:02}  \cite{MHM}.
The intersections of the various
curves in a very narrow range signals the crossover expected
at denaturation for this quantity and allows to locate the melting
temperature rather accurately, i.e. $\beta_c = 0.7525(3)$.
By fitting the initial slope of the critical pdf for chains up to $N=480$ 
we extrapolate $c=2.46(9)$, in good agreement with the exact enumeration 
results.

\begin{figure}[b]
\centerline{
\psfig{file=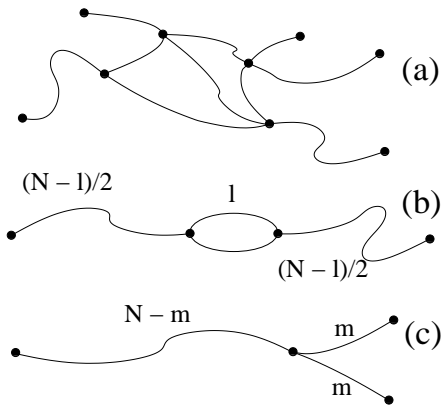,width=5.8cm}}
\vskip 0.2truecm
\caption{(a) Configuration of a polymer network
with ${\cal L}=2$, $n_1 =5$, $n_3 = 1$ and $n_4 =3$. 
(b) Loop of length $l$ embedded in a chain of length $N -l$ and
(c) chain of length $N-m$ with bifurcating end segments of length $m$.}
\label{fig:03}
\end{figure}  

The value of $c$ for loops with $l \ll N$ was predicted analytically
\cite{Kafr00} using exact results for entropic exponents of 
networks of arbitrary topology \cite{Dupl86,Ohno88}.
For a network of fixed topology $G$ (see example in Fig. \ref{fig:03}(a)), 
with $n$ self- and mutually avoiding segments, Duplantier \cite{Dupl86}, 
on the basis of renormalization group arguments, postulated the following 
scaling form for the total number of configurations:
\begin{equation}
\Gamma_{G} \sim \mu^{N} N^{\gamma_{G}-1} \ \ f \left( \frac{l_1}{N},
\frac{l_2}{N},.., \frac{l_n}{N} \right)
\label{Gamma}
\end{equation}
where $l_i$ is the length of the $i$-th segment, $N=\sum_i l_i$ and $f$
a scaling function. The value of $\gamma_G$ depends on
the number of independent loops, ${\cal L}$, and on the number of vertices
with $k$ legs, $n_k$, as
\begin{equation}
\gamma_G = 1 - {\cal L} d \nu + \sum_{k} n_k \sigma_k
\label{sigma}
\end{equation}
where $\nu$ is the radius of gyration exponent and $\sigma_k$,
$k=1,2,..$, are exactly  known exponents in $d=2$ \cite{Dupl86}:
\begin{equation}
\sigma_k = \frac{(2-k)(9k+2)}{64}\,.
\label{sigmak}
\end{equation}

This general scaling framework was applied to the DNA unzipping by 
considering relevant 
network topologies for the problem \cite{Kafr00,Kafr01}. For example, in
order to study the denaturated loop length pdf, one can assimilate the
situation of a typical loop within DNA to that of the loop in
Fig.~\ref{fig:03}(b). This amounts to assume that the action of the rest of 
the DNA molecule is the same as that of two long linear tails, thus totally
disregarding the presence of other loops. According to Eq.~(\ref{Gamma}), for
$G$ corresponding to the topology in Fig.~\ref{fig:03}(b), one has
\cite{Kafr00}
\begin{equation}
\Gamma_{\rm loop} \sim \mu^{2N} N^{-d\nu+2\sigma_1+2\sigma_3} h(l/N)\,.
\label{loop}
\end{equation}
In this equation we explicitly assume that, while the connectivity
of a loop step is equal to $\mu$, that of a step of the tails is
$\mu^{2}$. This assumption is consistent with the thermodynamics
of the denaturation transition and follows from the continuity
of the canonical free energy of the system, and from the fact
that each step of the tails corresponds in fact to two
steps of the loops. In this way the $l$ dependence of
the r.h.s. of Eq.~(\ref{loop}) does not enter in the
exponential growth factor, an important requisite for
the derivations below~\cite{notacop}.
For $N \gg l$ the number of configurations should reduce to that of a 
single double stranded
chain of length $N$, i.e. $\Gamma \sim \mu^{2N} N^{\gamma-1}$ with $\gamma = 
1+2\sigma_1$, according to Eq.~(\ref{Gamma}). This 
requires $h(x) \sim x^{-2\nu+2\sigma_3}$, for $x \ll 1$. 
So, the $l$ dependence in Eq.~(\ref{loop})
becomes $\sim l^{-c}$, with \cite{Kafr00} 
\begin{equation}
c = d \nu - 2 \sigma_3\,.
\label{exponentc}
\end{equation}
Clearly, in this approximation, $c$ is also the exponent by which the 
loop length pdf, $P$, scales. In reality, the segments departing from the 
two sides of the loop replace more complex fluctuating structures
containing denaturated bubbles of all sizes, separated by 
short linear double stranded segments. 
In $d=2$, $\nu =3/4$ and $\sigma_3 = -29/64$, therefore $c = 2 + 13/32 
\approx 2.41$, which is a value consistent within error bars with our 
numerical estimates.
As already known in $d=3$ \cite{Carl01}, this agreement implies that the
sequences of loops which ``dress" the two segments departing from the loop 
in Fig.~\ref{fig:03}(b) have very little effects on the value of $c$. 

We consider now the distribution of end-segments. With the
assumed boundary conditions, denaturated end segments of length $m$
occur in configurations where $\vec{r}_1 (N-m)=\vec{r}_2 (N-m)$ 
while $\vec{r}_1(k) \neq \vec{r}_2(k)$ for $k > N-m$.
The statistical geometry of denaturated end-segments is expected to be 
relevant for situations occurring in mechanical unzipping experiments
\cite{Kafr01}. Indeed, as a rule, this unzipping is induced by
applying forces which separate the strand extremes $\vec{r}_1(N)$
and $\vec{r}_2(N)$ by micromanipulation techniques \cite{micro}.
In the same spirit as in the case of denaturated loops, one can consider
now the network geometry of Fig.~\ref{fig:03}(c) \cite{Kafr01}, for which 
\begin{equation}
\Gamma_{\rm fork} \sim \mu^{2N} N^{3\sigma_1+\sigma_3} g(m/N)\,.
\label{fork}
\end{equation}
By matching again, for $m \ll N$, with the partition function of a SAW 
of length $N$, one gets for the distribution of the end segment
lengths $P_e(m,N) \sim m^{- \bar{c}}$ with \cite{Kafr01}
\begin{equation}
\bar{c} = - \left( \sigma_1 + \sigma_3 \right)\,.
\label{cbar}
\end{equation}

One should note that this last result holds also
in case one tries to match the behavior of $\Gamma_{\rm fork}$ with
that of a simple SAW of length $2N$ in
the limit $(N-m) \ll N$. In this case the
pdf results $\sim (N-m)^{-\bar{c}}$.

\begin{figure}[b]
\centerline{
\psfig{file=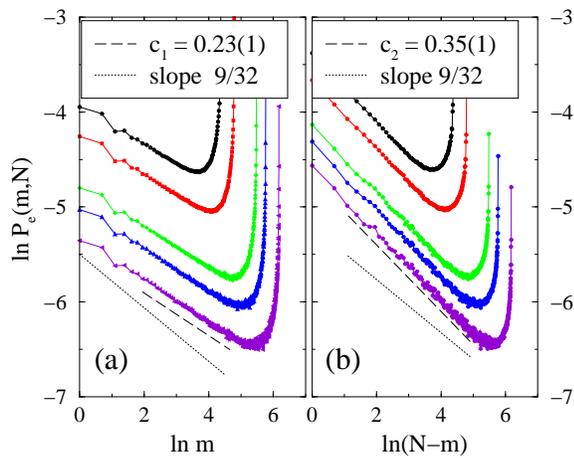,height=6.4cm}}
\vskip 0.2truecm
\caption{Plot of $\ln P_e(m,N)$ vs.\ $\ln m$ (a) and $\ln (N - m)$(b) at the 
estimated critical point $\beta_c = 0.7525$ and for $N = 80$, $120$, $240$, 
$320$ and $480$ (from top to bottom).}
\label{fig:04}
\end{figure}  

Figure \ref{fig:04} shows a plot of the logarithm of the pdf for
end segments of length $m$, $\ln P_e (m,N)$, as a function of (a) 
$\ln m$ and (b) $\ln (N - m)$. The data obtained from the PERM
show coexistence of two distinct power-law scaling behaviors, 
i.e.\ $P_e (m,N) \sim m^{-c_1}$ and $P_e (m,N) \sim (N - m)^{-c_2}$ where
a linear fit yields $c_1 = 0.23(1)$ and $c_2 = 0.35(1)$,
respectively. The slope determinations are rather sharp
in this case, and, in fact, the criterion of selecting
the denaturation temperature in correspondence with
the simultaneous manifestation of the power
law behaviors of $P_e$ is very efficient, and consistent with that
based on monitoring the behavior of
$\langle|\vec{r}_2(N)-\vec{r}_1(N)|\rangle$.

The existence of two distinct slopes, and the values of the exponents are 
in disagreement with what one expects on the basis of the network 
approximation, i.e. $\bar{c} = 9/32 \approx 0.28$ in $d=2$. 
Thus, for some reason, on $P_e(m,N)$ the effects of the structure of the 
double stranded part of the chain are noticeable and the schematization
through a simple polymer network topology is not fully adequate to represent 
the physics. In the case $N-m \ll N$ the number of single chain  
configurations is indeed that of a simple linear SAW of length $2N$, which is
asymptotically exactly known in two dimensions \cite{mu}.
On the other hand, for $m \ll N$, the 
configurations to count are those of an effective linear SAW chain of
length $N$, whose internal structure contains loops at all scales.
It is conceivable that not only the connective constants, but also
the entropic scaling properties of such
effective walk differ from those of a standard SAW. The difference
must concern the power law correction to the exponential growth
factor of the number of configurations as a function of $N$. Indeed,
as already remarked above,
the connectivity constant of the effective, double stranded walk 
at the transition, must be just the square of the simple SAW
connectivity constant. This follows from an obvious requirement of
free energy continuity and from the fact that in the high temperature
region the two strands behave as unbound simple SAW, with a total
length twice that of the effective double stranded walk, independent of
temperature \cite{notacarlo}.

To investigate this issue further we determined directly, on the basis of
PERM data, also the overall entropic behavior of the DNA chain. We 
considered two types of boundary conditions: 
(1) the extremes ($\vec{r}_1 (N)$ and $\vec{r}_2 (N)$) of the chain are 
free and (2) are forced to join in a single point ($\vec{r}_1 (N) = 
\vec{r}_2 (N)$), 
while in both cases, the strands have still a common 
origin ($\vec{r}_1(0)=\vec{r}_2(0)$).

Condition (1) is the one
applying to the effective walk discussed above.
We indicate with $Z_N^{(1)}$ and $Z_N^{(2)}$ the corresponding partition 
functions. In the spirit of the network approximation,
we neglect the contribution of denaturated loops within the double 
stranded phase one has $Z_N^{(2)} \sim \mu^{2N} N^{\gamma^{(2)} -1}$, 
with $\gamma^{(2)} = 1 + 2 \sigma_1$. 
We estimate also $Z_N^{(1)}$ within the same general framework,
by integrating the partition function $\Gamma_{\rm fork}$ of Eq.~(\ref{fork})
over all possible end-segment lengths. This integration gives

\begin{equation}
Z_N^{(1)} \sim \int_0^N dm \ \ \Gamma_{\rm fork} 
\sim \mu^{2N} N^{3 \sigma_1 + \sigma_3 +1},
\label{ZN1}
\end{equation}
as the integration of the scaling function $g(m/N)$ yields an
extra factor proportional to $N$.
Defining $Z_N^{(1)} \sim \mu^{2N} N^{\gamma^{(1)} -1}$ one eventually gets:
\begin{equation}
\gamma^{(1)} = 2 + 3\sigma_1 + \sigma_3\,.
\label{gammastar}
\end{equation}
This is a theoretical expression for the entropic
exponent of a DNA molecule as a whole, and can be 
directly compared
with numerical estimates, one of which already exists
in $d=3$ \cite{Caus00}, as we discuss in the next sections.

To calculate the entropic exponents $\gamma^{(1)}$ and $\gamma^{(2)}$
we estimated the quantity $(Z_{2N}^{(i)}/Z_N^{(i)})^{1/2N}$ 
by PERM sampling for reasonably long chains. For $N \to \infty$
one expects
\begin{equation}
\left( \frac{Z_{2N}^{(i)}}{Z_N^{(i)}} \right)^{\frac{1}{2 N}} \sim 
\mu \left( 1 + \ln 2 \ \frac{\gamma^{(i)} -1}{2N} \right),\ \ \ \ \ i=1,\,2\,.
\label{z2nzn}
\end{equation}

Since $\mu$ must coincide with the SAW connective constant, which in $d=2$ 
is very precisely known ($\mu = 2.63815852927(1)$ \cite{mu_square}), large 
$N$ data for the quantity on the l.h.s. of Eq. (\ref{z2nzn}) can be fitted 
by keeping as unique fitting parameter $\gamma^{(i)}$ in the r.h.s. 
expression. Alternatively, one can assume a value for $\gamma^{(i)}$ and check
whether data appear consistent with the assumed correction term
in the same r.h.s. expression. This consistency test is best 
applied to sets of data pertaining to different temperatures
close to the transition, as illustrated in the next section.
Since the correction term is rather sensitive to the 
choice of the temperature, this offers also a way of
locating the transition. Here we just mention that the slopes
predicted on the basis of the network approximation in
$d=2$ appear slightly, but definitely, inconsistent with the 
data, as discussed in the next section.

\section{Denaturating DNA as a copolymer network}

The polymer network representation of denaturating DNA \cite{Kafr00}
was originally proposed as a useful, but definitely approximate tool, 
without the possibility of controllable and systematic improvements.
For example, the environment seen by a single fluctuating loop
within the molecule was not proposed as something unique,
and choices slightly different from that discussed above
were also discussed \cite{Kafr00,Kafr01}. With these alternative choices,
leading to slightly different results,
the portions of the molecule surrounding the loop were
not necessarily treated as made of simple double segments.

Even if the approximate character remains,
all the discrepancies and inconsistencies 
discussed in the previous section can be resolved by a refinement of 
the whole picture and an improvement of the approach, which could 
allow more accurate predictions in the future. In the new perspective 
we propose, the rules
for associating a network schematization to loops or end-segments
should be regarded as unique. This follows from the fact that
the new polymeric entities one defines are supposed to
account, at an effective level, for all the complications
arising from the fluctuating geometry of the model.

The legs of the network are
either simple self-avoiding chains, in case they are really made of
single strands, or effective, dressed segments, when they represent 
fluctuating double stranded portions of the molecule
comprised between two given bound base pairs. This 
clear cut distinction 
suggests that the exponents $\sigma_k$, associated with the 
network vertices, should be 
modified with respect to the ``bare'' values considered so far, 
as soon as at least one outgoing 
leg is of the double stranded, dressed type.
Indeed, within the framework of a continuum Edwards model description
of the inhomogeneous polymer network, and of renormalization group ideas,
one can postulate the validity of homogeneity laws and
exponent relations analogous to Eqs. (\ref{Gamma}) and (\ref{sigma}),
with modified $\sigma$'s where appropriate. Something like this was 
already done
in Ref.~\cite{vonF97} where explicit field theoretical calculations
were performed in the case of copolymer star networks (no loops). 
These stars were composed of mutually
excluding branches made either of random walks, or of self avoiding 
walks. In the case considered here the copolymer is made of two kinds of
segments which do not differ as far as metric scaling exponents
are concerned (same $\nu$), but can have different entropic scaling
properties. It is indeed quite conceivable that an effective linear structure,
which should be resolved into a sequence of denaturated segments and loops
of all sizes, could have an entropic scaling different from
the one of a simple SAW on the lattice, even if the elongation
grows in the same way with $N$. 

\begin{figure}[b]
\centerline{
\psfig{file=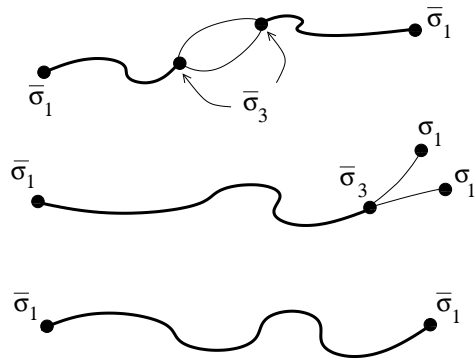,width=6.2cm}}
\vskip 0.2truecm
\caption{Representation of the geometries relevant for the DNA problem
by copolymer networks. Thick lines denote branches 
of the network composed by an alternating sequence of bound
segments and denaturated loops, while thin lines are genuine single
stranded segments. Only the latter have the same metric and
entropic scaling properties as a SAW.}
\label{fig:05}
\end{figure}  

\begin{figure}[b]
\centerline{
\psfig{file=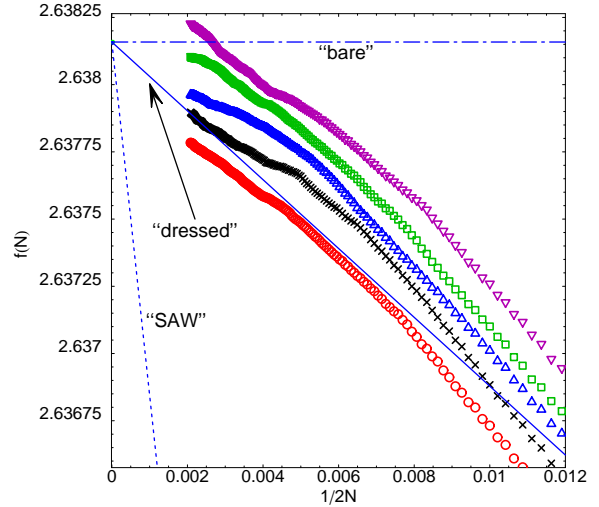,height=6.7cm}}
\vskip 0.5truecm
\caption{The function $f(N)$, defined in Eq.~(\ref{f_N}), for $N\to\infty$
should approach $\mu$ with zero slope if the $\gamma^{(1)}$ exponent would
be correctly determined by the homogeneous network approximation.
Instead, the trend predicted by the copolymer theory (continuous line)
fits rather 
well the data ($\beta=0.7520,\,0.7523,\,0.7526,\,0.7529,\,0.7532,$ from
below. We recall that we estimate $\beta_c=0.7525(3)$).}
\label{fig:06}
\end{figure}

The above considerations lead us to introduce two new exponents: 
$\overline{\sigma}_1 = \sigma_1 + \Delta \sigma_1$ for an isolated vertex 
with a dressed segment, and $\overline{\sigma}_3 = \sigma_3 + \Delta \sigma_3$ 
for a vertex joining a dressed segment and two self avoiding walks, as 
schematically illustrated in Fig.~\ref{fig:05}. As discussed above,
the introduction of such $\overline{\sigma}$'s amounts to
postulating a generalization of the scaling in Eq.~(\ref{Gamma})
to copolymer networks with two different types of segments.
Following the same arguments leading to Eqs.(\ref{exponentc}),
(\ref{cbar}) and (\ref{gammastar}) and using now the modified exponents 
wherever appropriate, one finds:
\begin{eqnarray}
c_1 &=& - \left( \sigma_1 + \sigma_3 \right) - \Delta \sigma_3 + 
\Delta \sigma_1\ ,
\label{eq1}
\\
c_2 &=& - \left( \sigma_1 + \sigma_3 \right) - \Delta \sigma_3 - 
\Delta \sigma_1\ , 
\label{eq2}
\\
c &=& d \nu - 2 \sigma_3 - 2 \Delta \sigma_3\ ,
\label{eq3}
\\
\gamma^{(2)} &=& 1 + 2 \sigma_1 + 2 \Delta \sigma_1\ ,
\label{eq4}
\\
\gamma^{(1)} &=& 2 + 3 \sigma_1 + \sigma_3 + \Delta \sigma_1 + 
\Delta \sigma_3\ ,
\label{eq5}
\end{eqnarray}
which is a set of five equations with only two unknown parameters,
$\Delta \sigma_1$ and $\Delta \sigma_2$.
These equations should be regarded as consistency requirements
in order to test the validity of the proposed copolymer picture.
Notice that within the copolymer network scheme the exponents associated
to the distributions of short and long end-segments are
distinct, $c_1 \neq c_2$, as soon as $\Delta \sigma_1 \neq 0$.
By solving the first two equations (\ref{eq1}) and (\ref{eq2}) with the 
numerical values for $c_1$ and $c_2$, we find $\Delta \sigma_3 = -0.01(1)$ 
and $\Delta \sigma_1 = -0.06(1)$. 
While the former is actually compatible with zero, the latter is not.
Once the values of $\Delta \sigma_1$ and $\Delta \sigma_3$ have been fixed, we 
can check for the consistency of the other exponents using Eqs. (\ref{eq3}),
(\ref{eq4}) and (\ref{eq5}).
Inserting the calculated $\Delta \sigma_3$
into Eq.~(\ref{eq3}) we find $c=2.43(2)$, i.e. a value slightly higher 
than what 
predicted from Eq.~(\ref{exponentc}). Indeed, our numerical estimates 
suggest for $c$ a slightly higher value than that predicted
on the basis of the bare network approximation, although the error bars 
cover both values.
Next we consider the last two equations (\ref{eq4}) and (\ref{eq5});
as both $\Delta \sigma_1$ and $\Delta \sigma_3$ are negative we expect that
$\gamma^{(1)}$ and $\gamma^{(2)}$ would be somewhat smaller than the values 
predicted from the homogeneous network approximation.
Substituting the above numerical values of $\Delta \sigma_1$ and 
$\Delta \sigma_3$ into Eqs.~(\ref{eq4}) and (\ref{eq5}) we find 
$\gamma^{(2)} \approx 1.22(2)$ and $\gamma^{(1)} \approx 1.99(2)$
(recall that the homogeneous network predictions are 
$\gamma^{(2)} = 1 + 11/32 \approx 1.34$ and $\gamma^{(1)} = 2 + 1/16 
\approx 2.06$).
We first examine case (1).
Since the differences between the ``bare'' and ``dressed'' values of 
$\gamma^{(1)}$ are small, to magnify the asymptotic details we 
considered the quantity
\begin{equation}
f(N)\equiv  \left(\frac{Z_{2N}^{(1)}}{Z_N^{(1)}} \right)^{\frac{1}{2 N}} -
\frac{(\gamma^{*} -1) \mu  \ln 2} {2 N}\ ,
\label{f_N}
\end{equation}
with $\gamma^{*} = 33/16$.
The coefficient of the $1/N$ term in the r.h.s.\ of the previous expression 
has been chosen such that if $Z_N^{(1)}$ would scale with the homogeneous
network exponent then $f(N)$ should approach the
connectivity constant $\mu$ with zero slope when plotted as function of
$1/2N$ (as follows from Eq.~(\ref{z2nzn})). 
A $\gamma^{(1)}$ larger (smaller) than its homogeneous network value
would imply a $f(N)$ approaching $\mu$ with a positive (negative) slope,
equal to $(\gamma^{(1)} -\gamma^*)\mu  \ln 2$.
Figure \ref{fig:06} shows a plot of $f(N)$ vs. $1/2N$ for five different 
temperatures around the estimated critical one. Clearly
$f(N)$ approaches $\mu$ in the limit of large $N$ with a negative 
slope. The solid line represents the slope for the value of 
$\gamma^{(1)}=1.99$ as obtained from Eq.~(\ref{eq5}), which apparently
fits reasonably well the data.

In the case of $Z_N^{(2)}$,
from all configurations generated by the PERM only those
where the two strands have common endpoints ($\vec{r}_1 (N) 
= \vec{r}_2 (N)$) are considered and this reduces considerably the statistics
available. 
The numerical results we find are consistent with both proposed pictures, 
since the low precision of data, due to the insufficient sampling, produces
large statistical errors.
So, the case (2), while also giving meaningful data, 
does not help in the determination of the right scenario.

We conclude that the improved representation based on the idea of a 
copolymer network with modified entropic exponents allows to match 
well all our numerical results, with just two adjustable parameters.

\section{The three dimensional case}
\label{sec:3d}

\begin{figure}[b]
\centerline{
\psfig{file=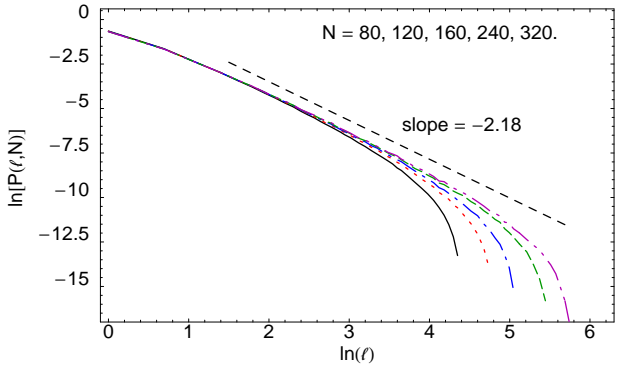,width=8.1cm}}
\vskip 0.5truecm
\caption{Log-log plot of the loops pdf at the critical point as function of 
their length for chains of various lengths. 
}
\label{fig:07}
\end{figure}

We now consider the $d=3$ case, where homogeneous network exponents 
can be deduced from estimates of $\gamma=1+2\sigma_1$~\cite{gamma3d}
($\sigma_1 \approx 0.079$)
or from $\varepsilon$-expansion results combined with resummation 
techniques ($\sigma_3 \approx -0.175$) \cite{Scha92}. 
From Eqs.(\ref{exponentc}) and (\ref{cbar}) one gets $c \approx 2.11$ and 
$\bar{c} \approx 0.095$. Previous
Monte Carlo simulations yielded $c=2.10(4)$ \cite{Carl01}. In the present
work we made an extra effort in order to get a reliable estimate
of the various exponents. With an analysis of the scaling behavior
of $\langle|\vec{r}_1(N)-\vec{r}_2(N)|\rangle$ we get here a very
precise estimate of the melting temperature [see Fig.~\ref{fig:02}(b)], 
consistent with that of Ref.~\cite{Caus00}. 
Extensive PERM sampling of the loop 
distribution at the estimated transition temperature yields
$c=2.18(6)$ (see Fig.~\ref{fig:07}). This refined value, confirming the first
order character of the transition, is slightly
higher than that reported in Ref.~\cite{Carl01}, although compatible
within the uncertainties. The discrepancy 
could be imputed to a slight overestimation of the transition
temperature made in that reference.
An analysis of the end-segments distribution yields, as in $d=2$, two 
slightly different exponents, $c_2 > c_1$, signalling again deviation 
from the prediction of the homogeneous network approximation. We estimate 
$c_2 = 0.16(1)$ and $c_1 = 0.14(1)$. These estimates are rather
sharp, and imply $\Delta \sigma_1=0.01(1)$ and $\Delta \sigma_3 = 0.055(10)$.
By inserting these values for the $\Delta \sigma$'s into our
expression for $c$, Eq.~(\ref{eq1}), we get $c \approx 2.22$,
which is well compatible with the last mentioned estimate
illustrated in Fig. \ref{fig:07}.

\begin{figure}[b]
\centerline{
\psfig{file=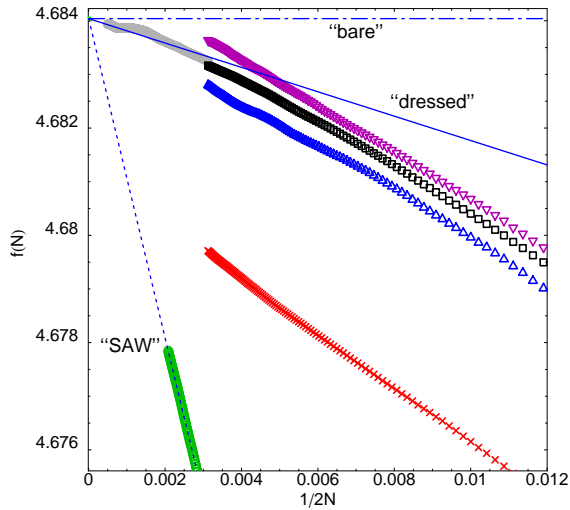,height=6.7cm}}
\vskip 0.5truecm
\caption{As in Fig.~\ref{fig:06}: from below, 
$\beta=0.293$ (SAW regime, evidenced by the dotted line, i.e.
$\gamma^*=\gamma$, with gamma representing the SAW exponent in
$d=3$ \protect\cite{gamma3d}), 
$1.335,\,1.3407,\,1.3413 (\beta_c),\,1.3419$.
The gray band shows data at $\beta_c$ and longer $N$ (up to 2000),
but with lower statistics.
}
\label{fig:08}
\end{figure} 

Causo {\it et al.} \cite{Caus00} determined in $d=3$ the exponent
$\gamma^{(1)} = 2.09(10)$, 
a value which should be compared with the Eqs. (\ref{gammastar}) and 
(\ref{eq5}) derived in this paper.
Inserting the above determined values for $\Delta\sigma_1$ and 
$\Delta\sigma_3$ in Eq.~(\ref{eq5}) we obtain $\gamma^{(1)} \approx 2.00$, 
to be compared with a value $\gamma^{(1)}\approx 2.06$ if one sticks to the 
homogeneous network prediction, given by Eq.~(\ref{gammastar}). As the 
$\gamma^{(1)}$ of Ref. \cite{Caus00} is compatible with both above values
we made an effort to estimate it again with improved accuracy.

Figure \ref{fig:08} shows a plot of $f(N)$ vs.~$1/2N$ for five different
temperatures of which three around $\beta_c$ (as for the $d=2$ case, we set
$\gamma^{*}$ equal to the value derived within the
homogeneous network picture for $\gamma^{(1)}$, $\gamma^{*} = 2.06$). 
The value of the connectivity constant is $\mu = 4.68404(9)$ for the cubic 
lattice \cite{mu_cubic}, a value indeed approached by all data sets for 
$\beta \leq \beta_c$. At high temperatures the data clearly show SAW
scaling, as expected, while at the transition point $f(N)$ seems indeed to
approach a slope given by the "dressed" exponent $\gamma^{(1)} = 2.00$ (solid
line). To confirm this, at the estimated critical point, we performed a series 
of calculations up to very long chains ($N = 2000$), but with somewhat lower
statistics. The latter data are plotted in gray in Fig. \ref{fig:08}:
indeed they seem to follow quite closely the solid line, as predicted by the 
inhomogeneous network theory.

Thus, even if the error bars on 
some exponents are still relatively large, the corrections suggested by
a block copolymer network representation of DNA seem to be
well consistent with the numerical scenario, like in the
two-dimensional case.

\section{Conclusion}

We investigated a lattice model of DNA denaturation both in $d=2$ and $d=3$ 
and determined the exponents associated to the decays of the pdf's for loops
and end-segments. For the loop pdf exponent we find $c >2$, which implies
a first order denaturation as the average loop length remains finite at the
transition point \cite{Pola66,Fish66}.
In $d=2$ the first order character is even more pronounced than in $d=3$ and
the values of $c$ are quite consistent with those predicted analytically on
the basis of entropic exponent of homogeneous networks \cite{Kafr00}.
An analysis of the end-segment
lengths reveals that the corresponding pdf displays two distinct
power law behaviors, one applying at small, and 
the other at large fork openings. This unexpected feature,
not predicted within the framework of the approximate description based 
on homogeneous polymer networks \cite{Kafr00}, suggests to 
describe
the DNA fluctuating geometry
as a copolymer network, in which a distinction is made between
single stranded and effective double stranded segments. 
The latter are assumed to have different entropic exponents than SAW's,
which lead to the introduction of a generalization of the previously known
expression for pdf and entropic exponents (see Eqs. (\ref{eq1}-\ref{eq5})).
We tested the compatibility of the observed results with this
effective copolymer network picture,
which is assumed to catch the essential physics of the
transition. Within this new framework, the different scaling
behaviors of the end-segment pdf can be qualitatively explained,
and the notion of overall entropic scaling of the macromolecule
acquires a precise and consistent meaning.
Especially in $d=2$, the numerical evidence that
a block copolymer network picture is compatible with the
overall data, is rather clear and convincing.

A main reason why the copolymer network description turns
out to be well compatible with the observed scalings is
probably the fact that the pdf of loop length scales always
with a sufficiently large $c$ exponent. So, even for $N \to \infty$, the
average width of the loops, and thus also of the dressed
segments, remains finite. This could be an important requisite
for the validity of the network picture. Indeed, it would be
interesting to test whether similar copolymer pictures
work also for other unzipping transitions of double
stranded polymers with a smaller $c$ \cite{Kafr01}. A natural
candidate is the unzipping occurring for the diblock
copolymer model studied in Ref. \cite{Baie01}. The physics of that
system can be assimilated to that of a DNA molecule in which
each of the two strands is made exclusively of one type of base,
and the bases of the two strands are complementary \cite{Kafr01}.
Of course, in such a model the loops forming can be made
with portions of different lengths of the two strands.
Moreover, different loops can also bind and form more
complicated topologies than in the DNA case.
By applying the simple homogeneous network picture to such a model, one 
would expect a second order transition with \cite{Kafr01}
$c = d \nu -2 \sigma_3 -1 \approx 1.4$ in $d=2$. The results for
the copolymer model of Ref. \cite{Baie01}, suggest a slightly, but
definitely higher value $c=1+9/16 \approx 1.56$, which is supported by
a connection with percolation theory. This discrepancy could be due to the 
more complicated topology of the loops and could indicate that the model is 
less favorable for the application of a network picture.

Homogeneous networks are very interesting {\it per se} and most recently 
were recognized as important tools also for the study of the topological 
entanglement of polymers \cite{Kard01}.
Block copolymer networks are a still relatively unexplored subject, 
in spite of the obvious fundamental and applicative interest. 
Thus, the realization that such networks could also be very relevant for a 
description of system like DNA at denaturation, at the moment, can not yield 
quantitative predictions based on field theory results.
Indeed, even the problem of identifying which kind of
copolymer model in the continuum, if any, could represent correctly
our discrete DNA in the scaling limit is far from trivial and 
completely open. For sure the identification of DNA as a
system potentially connected to the copolymer network physics
adds further interest to these intriguing objects, which
are already known to be somehow related to multifractal
aspects of polymer statistics \cite{vonF97} and, in the $d=2$ case, 
could realize interesting examples of conformal invariance.

We thank D. Mukamel for fruitful discussions and E. Orlandini
for ongoing collaboration. Financial support by MURST through COFIN
2001, INFM through PAIS 2001 and by European Network on "Fractal and 
Structures and Self-Organization" are gratefully acnkowledged.

\end{multicols}

\begin{references}
\bibitem{notacop} One should notice that already at this stage
the necessity of attributing different connective constants
to different segments confers a block-copolymer character
to the network. We will see below that the copolymer character
should naturally involve a more essential distinction of universal 
entropic exponents, in addition to that of connective constants.
\end{references}
\end{document}